\documentclass[useAMS,usenatbib]{mnras}

\usepackage{amsmath}
\usepackage{amssymb}
\usepackage{graphicx}
\usepackage{bm}
\usepackage{natbib}
\usepackage{color}
\usepackage{multirow}

\begin{document}
\title[DAMPE indicates a nearby slow-diffusion zone]{DAMPE proton spectrum indicates a slow-diffusion zone in the nearby ISM}

\author[K. Fang et al.]{
Kun Fang$^{1}$\thanks{fangkun@ihep.ac.cn}
Xiao-Jun Bi$^{1,2}$\thanks{bixj@ihep.ac.cn}
Peng-Fei Yin$^{1}$\thanks{yinpf@ihep.ac.cn}
\\
$^{1}$ Key Laboratory of Particle Astrophysics, Institute of High Energy 
Physics, Chinese Academy of Sciences, Beijing 100049, China\\
$^{2}$ School of Physical Sciences, University of Chinese Academy of Sciences, 
Beijing 100049, China\\
}

\maketitle

\begin{abstract}
The hardening and softening features in the DAMPE proton spectrum are very likely to be originated from a nearby supernova remnant (SNR). The proton spectrum from the nearby SNR is required to be very hard below $\approx10$ TeV. To reproduce this feature, we illustrate that anomalously slow-diffusion zone for cosmic rays (CRs) must be existed in the local interstellar medium (ISM) after also taking the dipole anisotropy of CRs into account. Assuming that the diffusion coefficient is homogeneous in the nearby ISM, we show that the diffusion coefficient is constrained to the magnitude of $10^{26}$ cm$^2$ s$^{-1}$ when normalized to 1 GeV, which is about 100 times smaller than the average value in the Galaxy. We further discuss the spatial distribution of the slow diffusion and find two distinct possibilities. In one case, the SNR is several hundred of parsecs away from the solar system, meanwhile both the SNR and the solar system are required to be included in a large slow-diffusion zone. The homogeneous diffusion belongs to this case. In the other case, the SNR is very close with a distance of $\sim50$ pc and the slow-diffusion zone is only limited around the SNR. The required diffusion coefficient is further smaller in the latter case. This work provides a new way of studying the CR diffusion in the local ISM.
\end{abstract}

\begin{keywords}
cosmic rays -- ISM: supernova remnants
\end{keywords}

\section{Introduction}
\label{sec:intro}
The cosmic-ray (CR) proton spectrum is not a single power-law function between GeV and PeV as predicted by the prevailing CR theory \citep[a recent review]{2019arXiv191101311L}. The spectral hardening at several hundred of GeV is almost a certain conclusion now, which has been confirmed by both the magnetic spectrometers \citep{2011Sci...332...69A,2015PhRvL.114q1103A} and the calorimeter detectors \citep{2009BRASP..73..564P,2019PhRvL.122r1102A,2019arXiv190912860A}. For the origin of the spectral hardening, there are different possible interpretations. The injection spectral indices could have a dispersion among the CR sources, which may account for the spectral hardening in the observed CR spectrum by the superposition effect \citep{2011PhRvD..84d3002Y}. The hardening may also be ascribed to the CR propagation. The diffusion coefficient for CR could be energy dependent due to, e.g., the different MHD properties among the Galactic interstellar medium (ISM) \citep{2014ApJ...782...36E}. Besides, the spectral fluctuation generated by nearby CR source(s) could alternatively explain the spectral hardening.

As the direct measurements of proton spectrum extend to higher energy, a second spectral break is discovered. The spectral softening at $\approx10$ TeV is confirmed by DAMPE with a significance of 4.7$\sigma$ \citep{2019arXiv190912860A} after being indicated by the earlier experiments \citep{2017ApJ...839....5Y,2018JETPL.108....5A}. This new spectral feature is not predicted by either the injection-revised model or the propagation-revised model mentioned above, while a discrete nearby CR source can naturally interpret both the spectral hardening and softening \citep{2019JCAP...10..010L,2019FrPhy..1524601Y}.

\begin{figure*}
 \centering
 \includegraphics[width=0.435\textwidth]{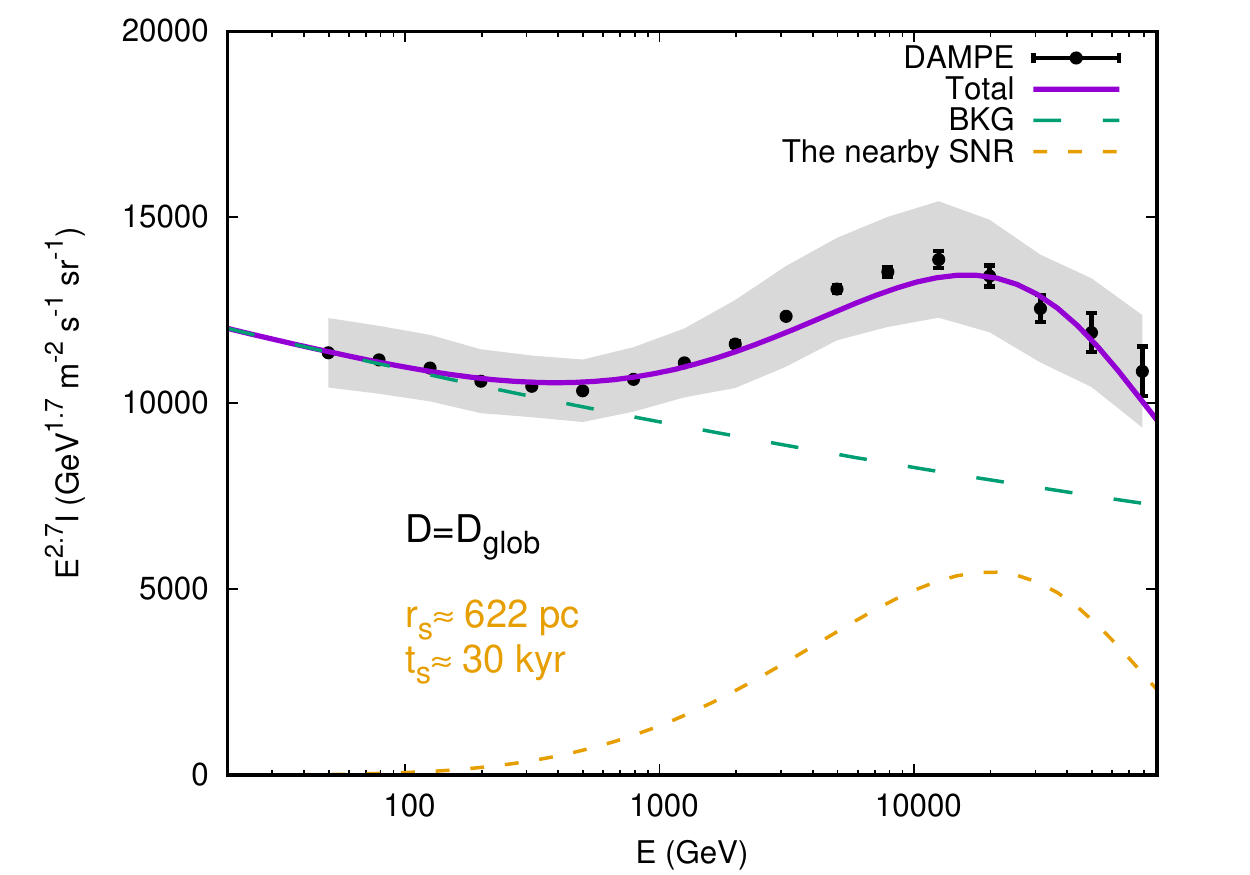}
 \includegraphics[width=0.40\textwidth]{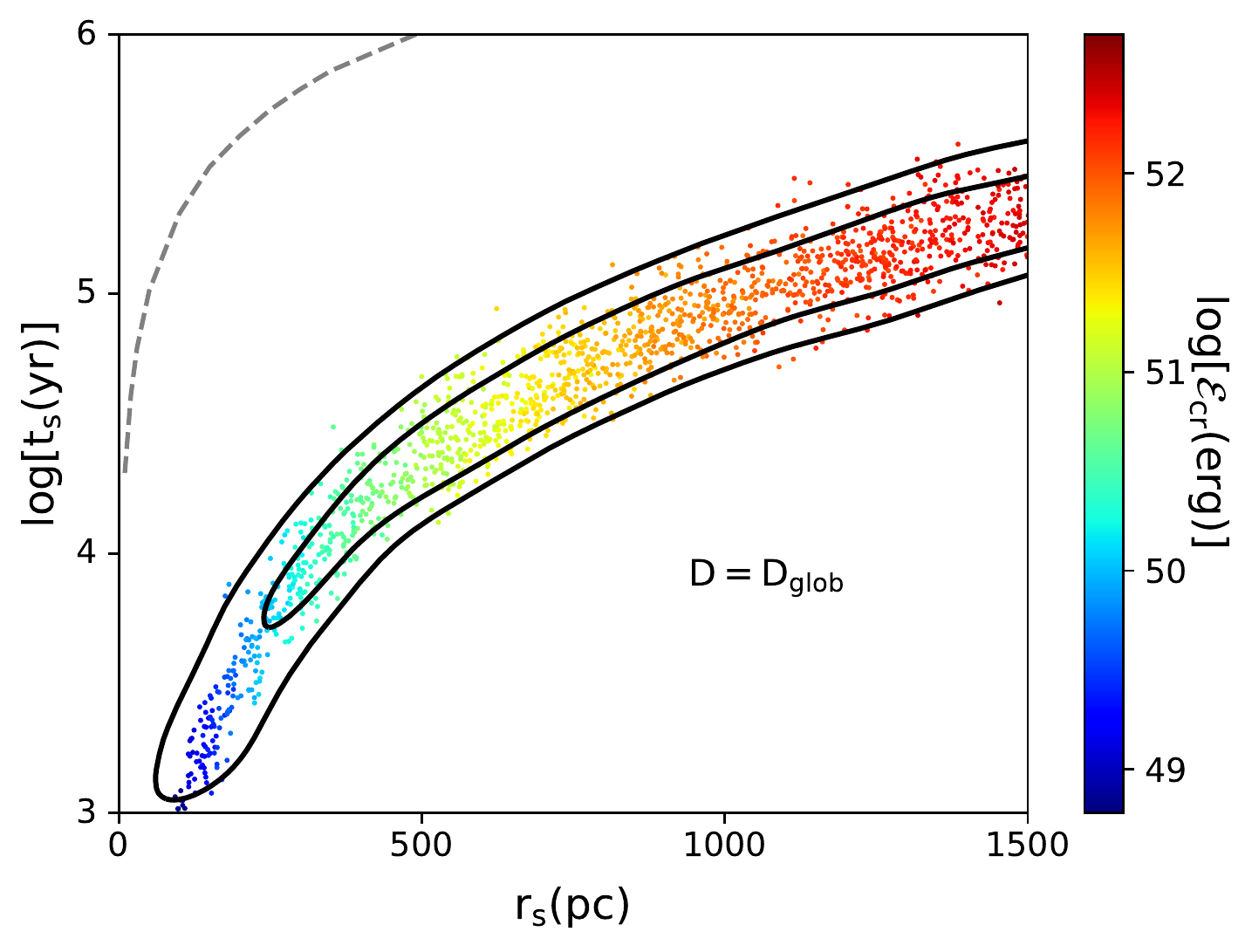}
 \caption{\textit{Left:} the proton spectrum calculated with the best-fit parameters toward the DAMPE data \citep{2019arXiv190912860A}. The best-fit parameters of the nearby SNR are: $r_s\simeq622$ pc, $t_s\simeq30$ kyr, $R_c\simeq53$ GV, and $\mathcal{E}_{cr}\simeq1.8\times10^{51}$ erg. The diffusion coefficient is set to be the default in GALPROP, which is the Galactic average value. \textit{Right:} the 3D posterior distribution of $r_s$, $t_s$, and $\mathcal{E}_{cr}$, corresponding to the scenario of the left figure. The gray dashed line is $3r_s/(2ct_s)=2.5\times10^{-3}$, which is roughly estimated from the measurements of CR dipole anisotropy. The posterior distributions in this work are all plotted with GetDist \citep{Lewis:2019xzd}.}
 \label{fig:test}
\end{figure*}

Moreover, there are intriguing correlations between the proton spectrum and the CR dipole anisotropy, which can be naturally comprehended in terms of the nearby-source explanation \citep{2019JCAP...10..010L}. The anisotropy amplitude increases from $\approx 1$ TeV to $\approx 10$ TeV and then decreases up to $\approx 100$ TeV, keeping pace with the proton spectral features. These could be commonly explained by a discrete CR source which remarkably contributes to the CR flux in 1-100 TeV. Meanwhile, the anisotropy phase suddenly switches from R.A.$\approx$4 hrs to the direction of the Galactic center at $\approx 100$ TeV, as the flux from the discrete source loses the dominance above $\approx 100$ TeV under the nearby-source scenario. In contrast, the energy dependence of the amplitude and phase of the anisotropy can hardly be explained by collective effects like the revised acceleration or propagation scenarios.

In this work, we explain the DAMPE proton spectral features with a nearby supernova remnant (SNR), as DAMPE is so far the only experiment that covers both the two spectral features \citep{2019arXiv191101311L}. Different from previous works, we concentrate on the implications on the CR diffusion in the nearby ISM. The HAWC $\gamma$-ray observations discover anomalously slow diffusion in the ISM around two nearby pulsars \citep{2017Sci...358..911A}. The derived diffusion coefficient is several hundred times smaller than the Galactic average, while the observations cannot decide whether the slow diffusion only happens around the pulsars or it is common in the nearby ISM. As the proton spectrum from the nearby SNR is only sensitive to the local diffusion coefficient rather than the global parameters, it is a good chance to investigate the local diffusion environment with the DAMPE spectrum.

In the next section, we will show that slow diffusion must be existed in the nearby ISM to reconcile the DAMPE proton spectrum and the dipole anisotropy of CRs. Then in Section \ref{sec:value}, we constrain the value of the diffusion coefficient assuming homogeneous slow diffusion in the local ISM. It is also possible that the slow-diffusion zone does not fill the whole nearby ISM, and we further discuss the spatial scale of the slow diffusion in Section \ref{sec:scale}. Finally, we conclude in Section \ref{sec:conclude}.

\section{Explain the DAMPE proton spectral features}
\label{sec:test}
The propagation of CR nuclei is dominated by the diffusion process for $E\gtrsim50$ GeV. For a nearby SNR, we can safely solve the diffusion equation in a spherical geometry with infinite boundary condition, as the scale of interest is much smaller than the vertical size of the diffusion halo \citep[several kpc,][]{2017PhRvD..95h3007Y}. We assume the SNR to be a point source with burst-like CR injection, then the differential intensity at the Earth is written as 
\begin{equation}
 I(E,r_s,t_s)=\frac{c}{(4\pi)^{5/2}(Dt_s)^{3/2}}{\rm exp}\left(-\frac{r_s^2}{4Dt_s}\right)Q(E)\,,
 \label{eq:diff}
\end{equation}
where $c$ is the speed of light, $r_s$ and $t_s$ are the distance and age of the SNR respectively, $D(E)=D_0(R/{\rm 1\,GV})^\delta$ is the diffusion coefficient, and $R=E/(Ze)$ is the rigidity of nucleus with charge number $Z$. The injection energy spectrum is assumed to be $Q(E)=Q_0(E/{\rm 1\,GeV)}^{-\gamma}{\rm exp}(-E/E_c)$, where $E_c$ is decided by a cutoff rigidity $R_c$ since the acceleration only depends on the rigidity of the nucleus. The spectral softening at $\approx10$ TeV could be ascribed to the injection cutoff.

At this moment, we superpose the spectrum of the nearby SNR on a single power-law background to fit the DAMPE data. The spectral hardening appears at $\approx 500$ GeV in the DAMPE measurement \citep{2019arXiv190912860A}, below which we assume the nearby SNR contributes little to the spectrum. Using the data from $\approx50$ GeV to $\approx200$ GeV, we get the backgound spectrum with the form of $I_0(E/{\rm 100\,GeV})^{-\alpha}$, where $I_0=4.34\times10^{-2}$ GeV$^{-1}$ m$^{-2}$ s$^{-1}$ sr$^{-1}$ and $\alpha=2.76$.

As the spectral hardening is significant, the spectrum from the nearby SNR is required to be very hard below $\approx10$ TeV. Obviously, the hard spectrum could have several origins: i) the second term (exponential term) in equation (\ref{eq:diff}) dominates, which means most of the low-energy protons have not arrived at the Earth yet. This could be due to the large distance or the young age of the SNR; ii) for the same reason, a small diffusion coefficient could account for the hard spectrum; iii) it could be alternatively originated from a hard injection spectrum, i.e., a small $\gamma$. 

\begin{figure*}
 \centering
 \includegraphics[width=0.40\textwidth]{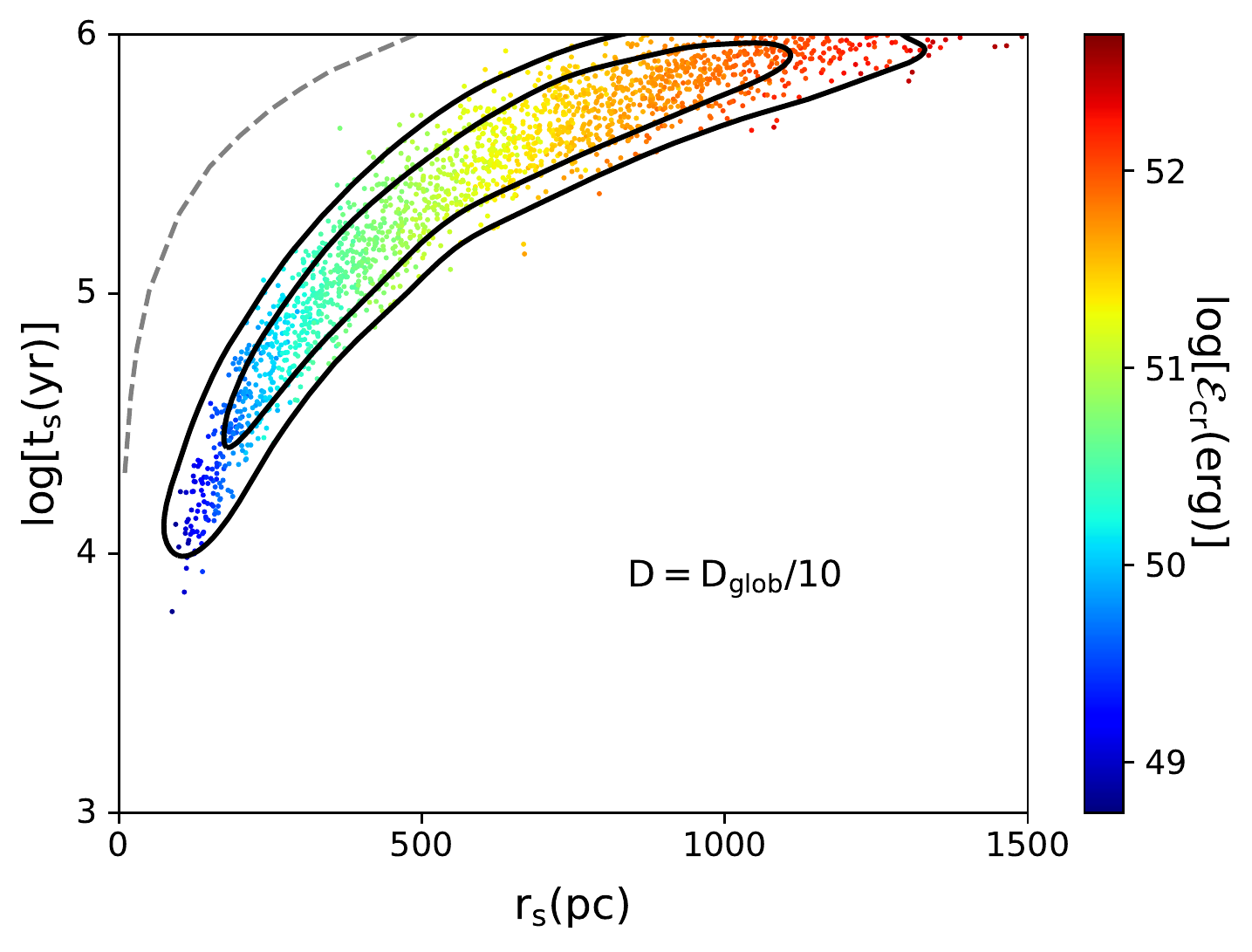}
 \includegraphics[width=0.40\textwidth]{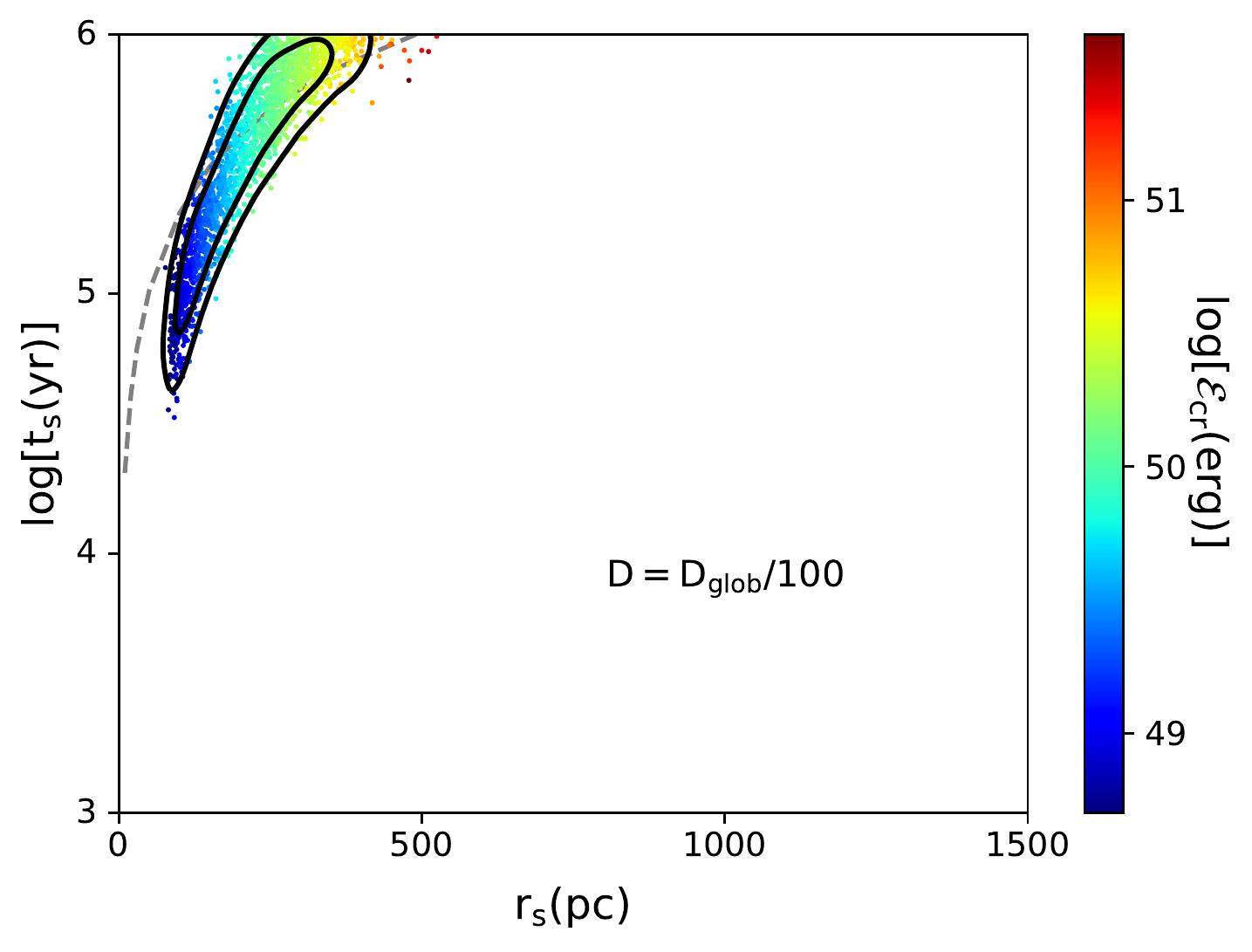}
 \caption{Same as the right graph of Fig. \ref{fig:test}, while here we adopt smaller diffusion coefficients ($D=D_{\rm glob}/10$ in the left panel and $D=D_{\rm glob}/100$ in the right panel).}
 \label{fig:test2}
\end{figure*}

Firstly, we vary the distance and age of the SNR, while fix the diffusion coefficient to be the global average in the Galaxy, which is denoted by $D_{\rm glob}$. We take the default in GALPROP\footnote{https://galprop.stanford.edu/} as $D_{\rm glob}$, i.e., $D_0=3.86\times10^{28}$ cm$^2$ s$^{-1}$ and $\delta=0.33$, which is roughly consistent with the results derived from the boron-to-carbon (B/C) ratio \citep[e.g.,][]{2017PhRvD..95h3007Y}. For the injection spectral index we take $\gamma=2.15$, which is common for models involving Fermi acceleration \citep{2001JPhG...27..941E}. Then the free parameters are: $r_s$, $t_s$, $R_c$, and $\mathcal{E}_{cr}$. The last parameter is the energy injected to protons, which determines the normalization $Q_0$ of the injection spectrum.

We fit the DAMPE proton spectrum considering both the statistical and systematic uncertainties being added in quadrature, while we neglect the correlations in the systematic errors. We apply the MULTINEST software \citep{2008MNRAS.384..449F,2009MNRAS.398.1601F,2019OJAp....2E..10F} to perform the posterior inferences. The spectrum with the maximum likelihood parameters is shown in the left of Fig. \ref{fig:test}, compared with the DAMPE data. Obviously, there is strong degeneracy between $r_s$ and $t_s$ as shown in the right of Fig \ref{fig:test}. The distribution of $\mathcal{E}_{cr}$ is also shown in right with color bar. 

Meanwhile, $r_s/t_s$ is also constrained by the dipole anisotropy of CRs. The dipole anisotropy of an isolate source is $3r_s/(2ct_s)$ \citep{1971ApL.....9..169S}. The current measurements of dipole anisotropy are roughly consistent with each other \citep[see Fig. 6 in][]{2017PrPNP..94..184A}. The anisotropy above $\approx100$ TeV has a power-law form, which can be well explained by the CR background. Below $\approx100$ TeV, the anisotropy could be dominated by the nearby SNR. Here we only give a rough estimation for the anisotropy; we leave the more careful calculation in the next section. The anisotropy peak below $100$ TeV is $\approx10^{-3}$, which corresponds to the CR flux peak of the nearby SNR. As shown in Fig. \ref{fig:test}, the nearby SNR has a largest contribution of $\approx1/2.5$ of the total flux. So we have $3r_s/(2ct_s)\approx2.5\times10^{-3}$ (see equation (\ref{eq:ani})) and show this relation in the right of Fig. \ref{fig:test} with a dotted line. The $r_s$ to $t_s$ ratio derived by the proton spectral fitting is much larger than the value required by the anisotropy measurement. This indicates that we cannot explain the proton spectrum and the anisotropy simultaneously by simply varying the distance and age of the nearby SNR.

Then we adopt smaller diffusion coefficients with $D=D_{\rm glob}/10$ and $D=D_{\rm glob}/100$ respectively, and do the fitting again. The 3D posterior distributions are shown in Fig. \ref{fig:test2}. We can see that only when the diffusion coefficient is small enough with $D=D_{\rm glob}/100$ can the contour overlap with the line required by the anisotropy. The anisotropy of the nearby SNR is independent of $D$ under the diffusion process, thus we can reconcile the DAMPE spectral feature and the CR anisotropy by significantly reducing the diffusion coefficient.

Next we examine if the hard spectrum from the nearby SNR can be purely explained by a hard injection spectrum. We set $\gamma$ as a free parameter and perform the fitting procedure again, while we add a constraint of $3r_s/(2ct_s)=2.5\times10^{-3}$ to ensure the consistency with the anisotropy measurement. The fitting result shows that the injection spectrum is required to be $\gamma\simeq1.53\pm0.13$. This is not a reasonable value, which is much harder than that indicated by either theories or observations. 

All the analyses above illustrate that a slow-diffusion zone with $D$ significantly smaller than the Galactic average must be existed in the nearby ISM. 

\section{Constrain the local diffusion coefficient}
\label{sec:value}
In this section, we give a more careful calculation to constrain the value of the diffusion coefficient in the local environment. We also include the AMS-02 proton data \citep{2015PhRvL.114q1103A} in the fitting procedure. The proton spectrum of AMS-02 is so far the most accurate measurement below 1 TeV. The DAMPE result is in agreement with the AMS-02 result after considering the systematic errors. The AMS-02 spectral index is stable between $\approx50$ GeV and $\approx200$ GeV \citep{2015PhRvL.114q1103A}, which is not affected by the solar modulation in lower energies and the spectral hardening in higher energies. So the AMS-02 data in this energy range can be used to determine the background spectrum. 

\begin{figure*}
 \centering
 \includegraphics[width=0.45\textwidth]{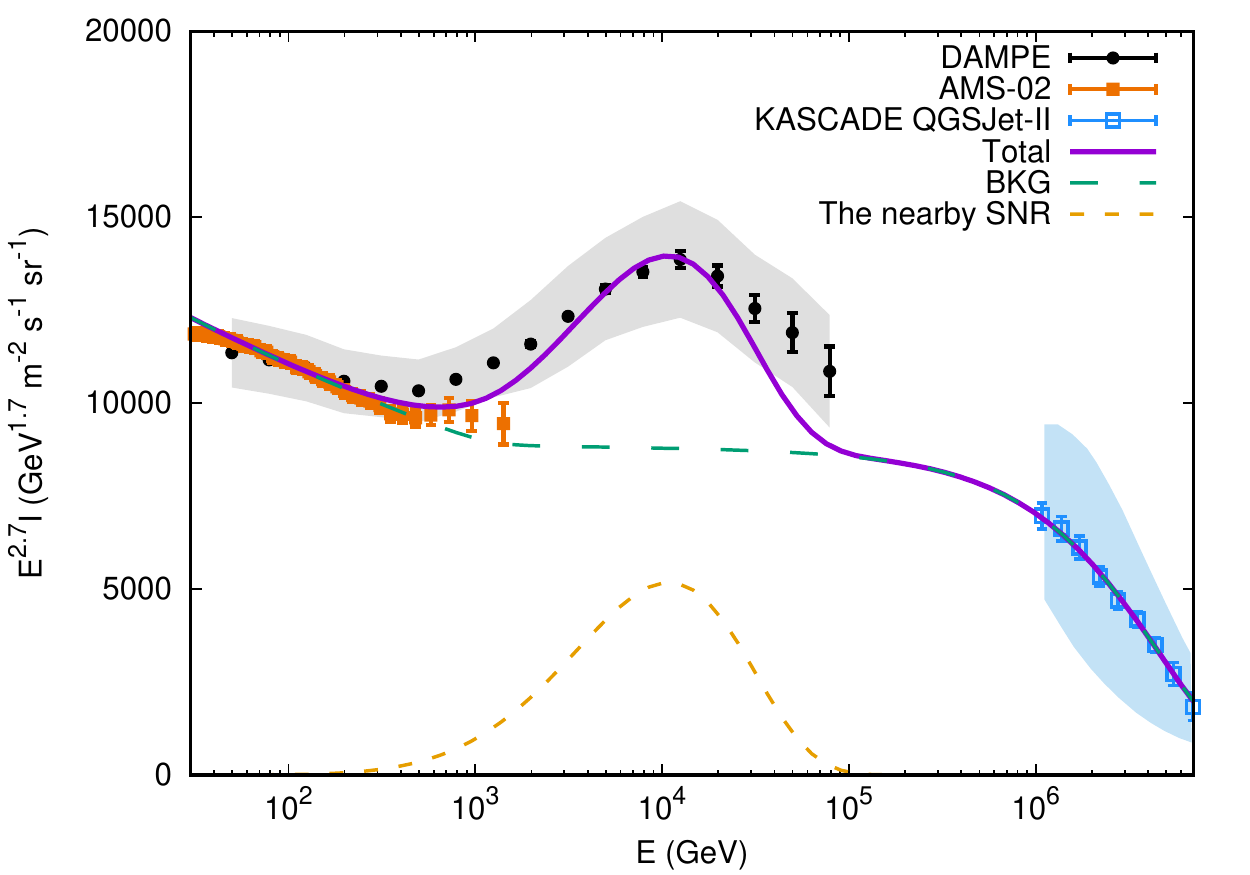}
 \includegraphics[width=0.45\textwidth]{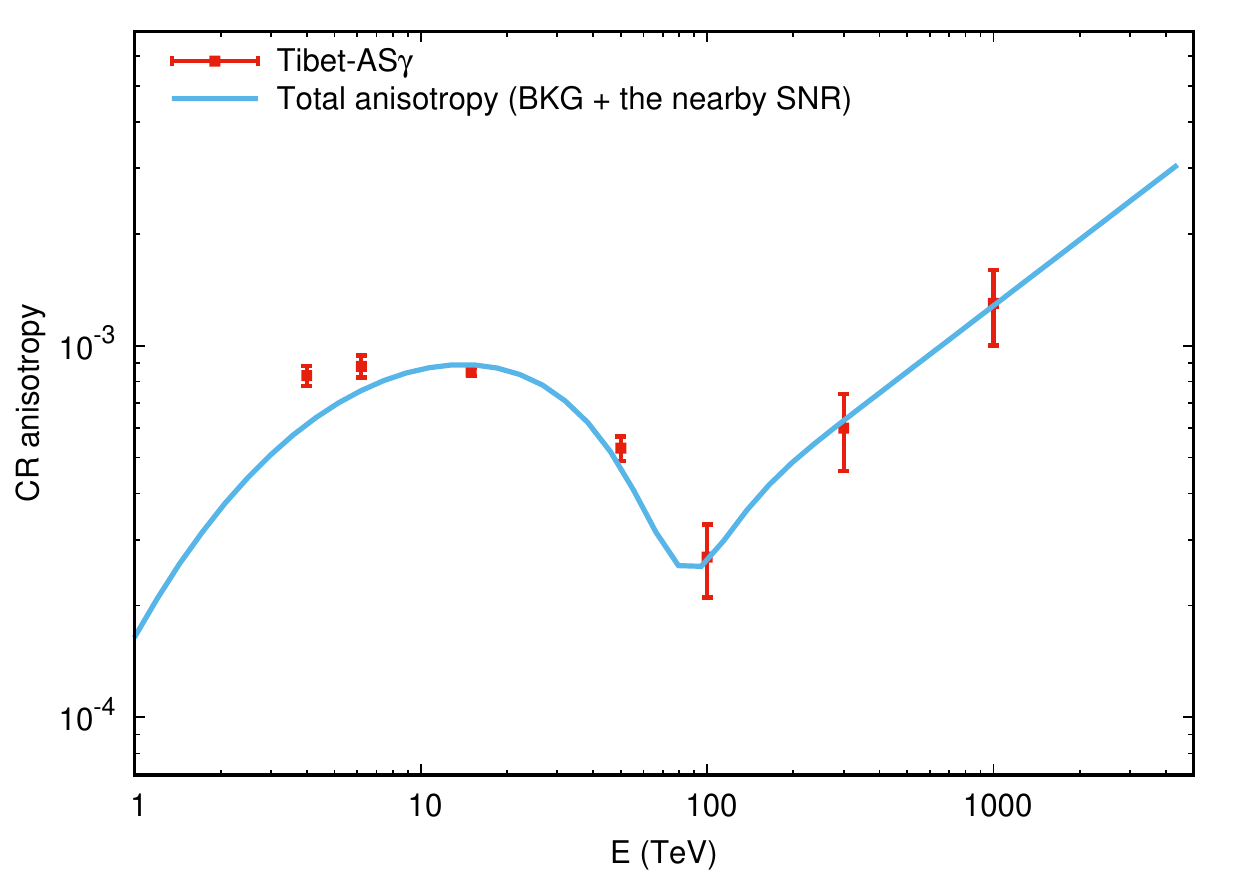}
 \caption{\textit{Left}: the proton spectrum calculated by the best-fit parameters toward both the proton data \citep{2015PhRvL.114q1103A,2019arXiv190912860A} and the CR anisotropy data \citep{2005ApJ...626L..29A,2015ICRC...34..355A}. We fix the source distance in the fitting procedures and test three different distances: $r_s$=100, 250, and 500 pc. The result shown here is the case of $r_s=250$ pc. See Table \ref{tab:result} for the best-fit parameters of all the cases. \textit{Right}: the best-fit CR anisotropy for the same scenario with the left graph.}
 \label{fig:r250}
\end{figure*}

\begin{table*}
 \centering
 \caption{Fitting results to the DAMPE and AMS-02 proton spectra and the dipole CR anisotropy of Tibet-AS$\gamma$}
 \begin{tabular}{ccccc}
  \hline
  \hline
  \multicolumn{2}{c}{} & $r_s=100$ pc & $r_s=250$ pc & $r_s=500$ pc \\
  \hline
  \multirow{2}*{$D_0$ ($10^{26}$ cm$^2$ s$^{-1}$)} & Best fit & 1.00 & 2.54 & 5.12 \\
   & $1\sigma$ range & [0.89,1.32] & [2.18,3.30] & [4.76,7.09] \\
  \multirow{2}*{$t_s$ (100 kyr)} & Best fit & 2.00 & 4.92 & 9.84 \\
   & $1\sigma$ range & [1.70,2.13] & [4.24,5.40] & [8.17,10.3] \\
  \multirow{2}*{$R_c$ (TV)} & Best fit & 18 & 17 & 17 \\
   & $1\sigma$ range & [16,20] & [16,20] & [16,20] \\
  \multirow{2}*{$\mathcal{E}_{\rm cr}$ ($10^{50}$ erg)} & Best fit & 0.16 & 2.5 & 20 \\
   & $1\sigma$ range & [0.12,0.18] & [1.8,3.0] & [14,21] \\
  \multirow{2}*{$c_1$ ($10^{-3}$)} & Best fit & 1.28 & 1.28 & 1.32 \\
   & $1\sigma$ range & [1.12,1.60] & [1.09,1.59] & [1.08,1.52] \\
  \multirow{2}*{$c_2$} & Best fit & 0.61 & 0.59 & 0.62 \\
   & $1\sigma$ range & [0.54,0.80] & [0.52,0.80] & [0.53,0.77] \\
  \hline  
 \end{tabular}
 \label{tab:result}
\end{table*}

However, the extensive air showers (EAS) experiments working at higher energies indicate that the background spectrum may not be described by a single power law. We adopt the latest KASCADE result \citep{2013arXiv1306.6283T}, which uses QGSJET-II to model the hadronic interactions, to determine the high-energy end of the proton background. A smoothly broken power-law function is enough to accommodate the AMS-02 and the KASCADE data, which has the form of $I_0(E/{\rm 100\,GeV})^{-\alpha}[1+(E/E_b)^s]^{\Delta\alpha/s}{\rm exp}(-E/E_c)$. We fix the smoothness of the break $s$ to be 5.0 like the DAMPE paper, which affects little to the result in terms of the present data \citep{2019arXiv190912860A}. The best-fit parameters toward the AMS-02 data in 50-200 GeV and the KASCADE data are: $I_0\simeq4.40\times10^{-2}$ GeV$^{-1}$ m$^{-2}$ s$^{-1}$ sr$^{-1}$, $\alpha\simeq-2.789$, $E_b\simeq1.19$ TeV, $\Delta\alpha\simeq0.086$, and $E_c\simeq4.73$ PeV.

Different from the rough estimation in the previous section, we include the energy-dependent  anisotropy amplitude in the following fitting procedures. We adopt the large-scale dipole anisotropy measured by Tibet-AS$\gamma$ \citep{2005ApJ...626L..29A,2015ICRC...34..355A}. It covers the most anisotropy features among the present experiments. According to the definition, the total dipole anisotropy is 
\begin{equation}
 \Delta=\frac{\sum_i\bar{I}_i\Delta_i\bm{n}_i\cdot\bm{n}_{\rm max}}{\sum_{i}\bar{I}_i}\,,
 \label{eq:ani}
\end{equation}
as derived by \citet{1971ApL.....9..169S}. $\bar{I}_i$ is the mean CR intensity of different components. We denote the intensity of the nearby SNR as $\bar{I}_{\rm ns}$, the intensity of the background as $\bar{I}_{\rm bg}$. The anisotropy from the nearby SNR $\Delta_{\rm ns}$ is decided by the distance and the age of the source as described above. The anisotropy from the CR background $\Delta_{\rm bg}$ should be a power-law form \citep{2017PrPNP..94..184A}. We assume $\Delta_{\rm bg}=c_1(E/{\rm 1\,PeV})^{c_2}$ in the following fitting procedures, where $c_1$ and $c_2$ are set to be free. The location of the nearby SNR $\bm{n}_{\rm ns}$ is set to be RA=4$^h$0$^m$, Dec=$-24^\circ$30' to accommodate the anisotropy direction below $\approx100$ TeV \citep{2019JCAP...10..010L}, while the direction of the background $\bm{n}_{\rm bg}$ points to the Galactic center. $\bm{n}_{\rm max}$ is the direction of maximum CR intensity. 

The anisotropy below $\approx100$ TeV is dominated by the nearby SNR, while the CR background dominates the anisotropy above $\approx100$ TeV. So the CR elemental composition only affects our results below $\approx100$ TeV. As the CR flux is mainly contributed by proton and helium for $\lesssim100$ TeV, we can safely neglect heavier nuclei in the calculation of $\bar{I}_i$. For the nearby SNR, we assume the helium injection spectrum has the same $\gamma$ and $R_c$ with that of proton, while we determine $Q_0$ according to the ratio of abundance between proton and helium provided by GALPROP. We find that the ratio of injection energy between proton and helium is about 2:1, consistent with \citet{2019JCAP...10..010L}. For the helium background, we adopt the function provided by AMS-02 before the spectral hardening: $I_0(E/{\rm 100\,GeV})^{-\alpha}$, where $I_0=3.54\times10^{-2}$ GeV$^{-1}$ m$^{-2}$ s$^{-1}$ sr$^{-1}$ and $\alpha=2.78$.

\begin{figure*}
 \centering
 \includegraphics[width=0.45\textwidth]{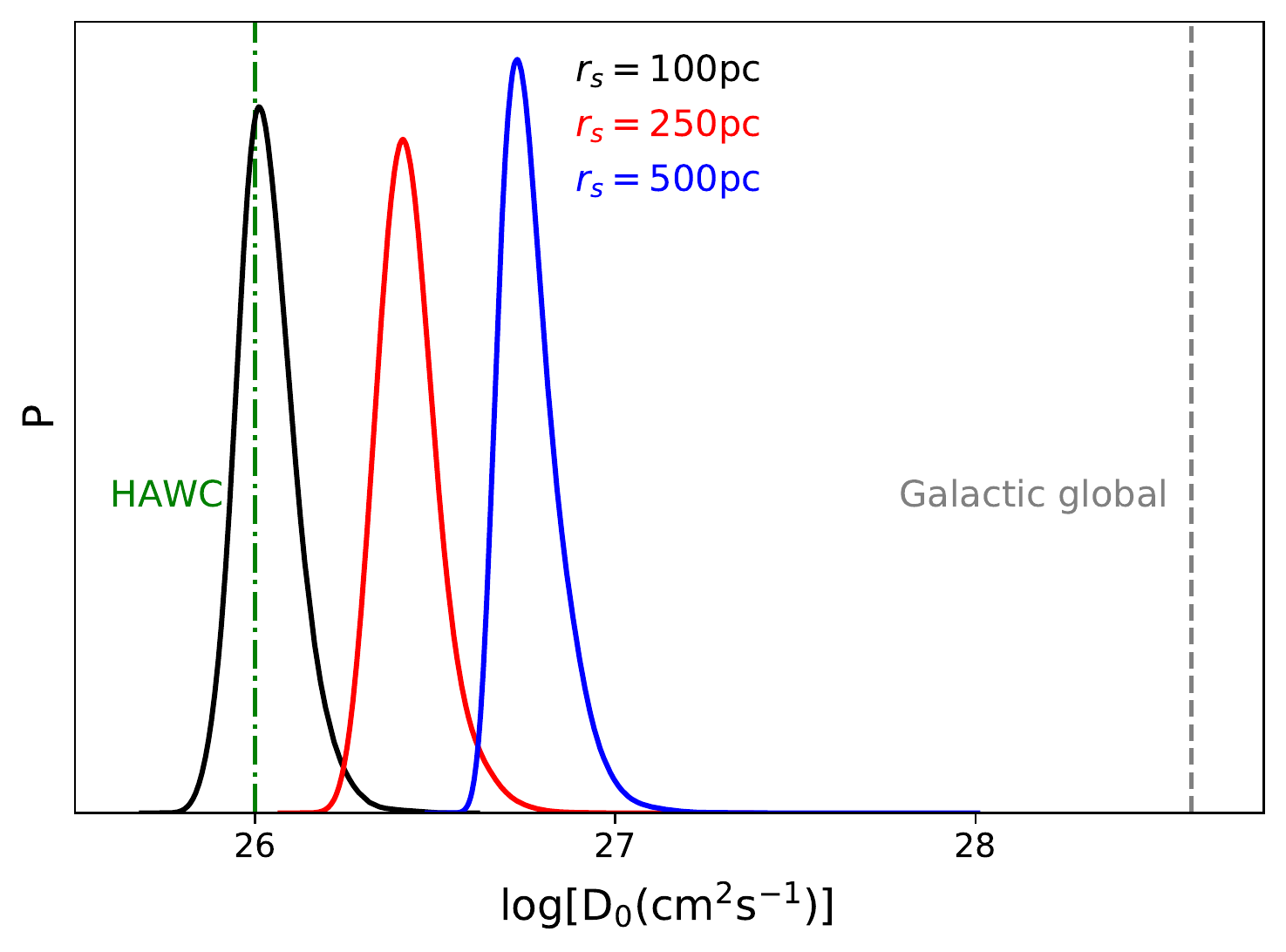}
 \includegraphics[width=0.45\textwidth]{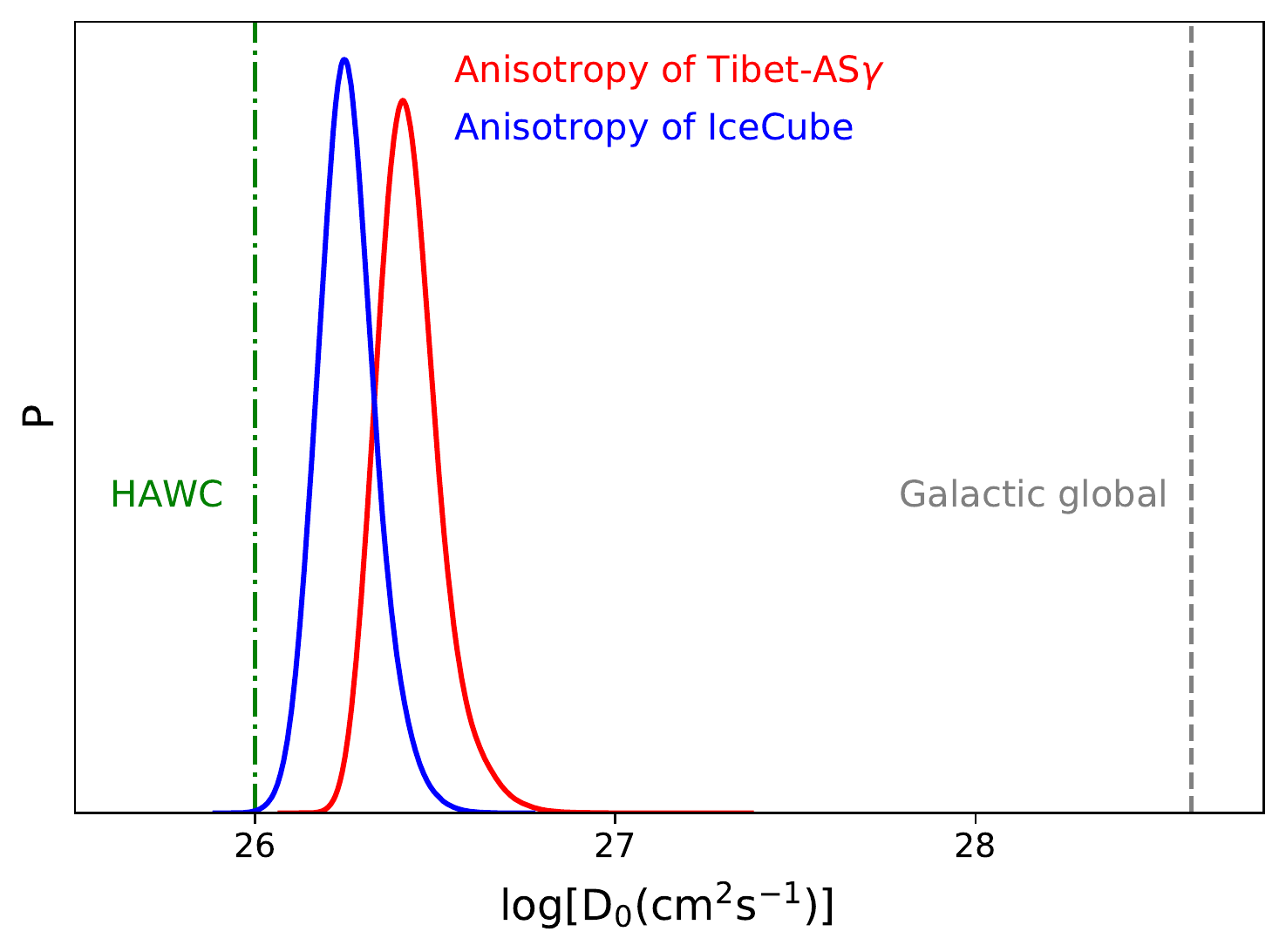}
 \caption{1D posterior distributions of $D_0$, which are obtained by fitting to the proton data and the CR anisotropy simultaneously. Since there is degeneracy between the parameters (see the text for detail), we fix the distance of the nearby SNR in each fitting procedure. \textit{Left}: we show the cases with different SNR distance $r_s$, compared with the Galactic average value derived by the boron-to-carbon ratio and the value in the circumstance of two nearby pulsars measured by the $\gamma$-ray observations of HAWC \citep{2017Sci...358..911A}. The CR dipole anisotropy measured by Tibet-AS$\gamma$ is adopted \citep{2005ApJ...626L..29A,2015ICRC...34..355A}. \textit{Right}: we show the results obtained with different anisotropy measurements \citep{2016ApJ...826..220A}, while for both the cases we fix $r_s$ to be 250 pc.}
 \label{fig:d0}
\end{figure*}

In terms of the data fitting, there is still significant degeneracy between the parameters (especially among $r_s$, $t_s$, and $\mathcal{E}_{\rm cr}$). This can be comprehended through the right panel of Fig. \ref{fig:test2}, as the dotted line is intercepted by the contour for quite a wide range. So we set the source distance to be constant in the fitting procedures and test the cases of $r_s=100$ pc, 250 pc, and 500 pc by performing three different groups of fitting. The injection spectral index is still set to be 2.15 as predicted by the acceleration theory. Then the free parameters are: $t_s$, $R_c$, $\mathcal{E}_{\rm cr}$, $D_0$, $c_1$, and $c_2$.

The results of parameter fitting are presented in Table \ref{tab:result}, while the best-fit proton spectrum and CR anisotropy for the case of $r_s=250$ pc are shown in Fig. \ref{fig:r250} as an example. The required energy injected to CR is positively correlated with the source distance. For a typical SNR explosion energy of $10^{51}$ erg and a conversion efficiency of $\sim10\%$ for CRs, it is reasonable to have $\mathcal{E}_{\rm cr}\sim10^{50}$ erg. As shown in Table \ref{tab:result}, the required CR energies are about 0.2, 2.0, and 20 times of the typical value for the source distances of 100, 250, and 500 pc, respectively. This implies that the SNR distance should be in the range of 100-500 pc, otherwise the CR energy will be deviated too much from the typical value. 

The posterior distribution of $D_0$ is presented in the left of Fig. \ref{fig:d0}. $D_0$ is also positively correlated with $r_s$. For $100\,{\rm pc}<r_s<500\,{\rm pc}$, $D_0$ is constrained in the magnitude of $10^{26}$ cm$^{-2}$ s$^{-1}$, which is about 100 times smaller than the Galactic average value derived by B/C. Meanwhile, the diffusion coefficient measured by HAWC is $4.5\times10^{27}$ cm$^2$ s$^{-1}$ at $\sim100$ TeV for $\delta=0.33$ \citep{2017Sci...358..911A}. If we extrapolate the HAWC diffusion coefficient to 1 GeV, it should be $\approx10^{26}$ cm$^2$ s$^{-1}$. Thus, the diffusion coefficient in the nearby ISM derived by the CR measurements is coincident with those around the two nearby pulsars derived by the $\gamma$-ray observations. The slow diffusion of CRs is likely to be common in the local environment. We will discuss the spatial scale of the slow diffusion in the next section.

Besides, the age of the nearby SNR is required to be several hundred of kilo years, which can hardly be identified at the present day. This is coincident with the fact that no promising nearby CR source has been detected in the direction indicated by the CR dipole anisotropy. For an age of $\approx500$ kyr, the shock velocity of the SNR decreases to $\approx50$ km s$^{-1}$ for typical input parameters \citep{2017AJ....153..239L}, which is too slow to accelerate high-energy CRs. So the SNR may no longer be observable in high energies. 

We then discuss some potential uncertainties in the calculation. We take $\gamma=2.15$ as the injection spectra index in the above calculations, which is supported by the Fermi acceleration models \citep{2001JPhG...27..941E}. Indeed, both the $\gamma$-ray and radio observations of SNRs suggest a spectra slope steeper than 2 \citep{2011JCAP...05..026C,2011MNRAS.418.1208B}. We also note that the required slope of the background anisotropy is $\approx0.6$ ($c_2$ in Table \ref{tab:result}). As the energy dependence of the background anisotropy is determined by the energy dependence of the diffusion coefficient \citep[e.g.,][]{2017PrPNP..94..184A}, the energy index of the diffusion coefficient above $\approx100$ TeV should also be $\approx0.6$. If we extrapolate this index to GeV energies and consider the proton spectra slope of $\approx2.75$ before the hardening, it should be reasonable to assume $\gamma=2.75-0.6=2.15$. Besides, if the ISM turbulence is Kolmogorov type with $\delta=0.33$, the injection slope would be even steeper with $\gamma\approx2.4$. We perform the fitting procedures again with $\gamma=2.4$ and find this change impacts little on the diffusion coefficient. The reason is that the required hard spectrum of the nearby SNR is dominated by the exponential term in equation (\ref{eq:diff}). 

\begin{figure*}
 \centering
 \includegraphics[width=0.45\textwidth]{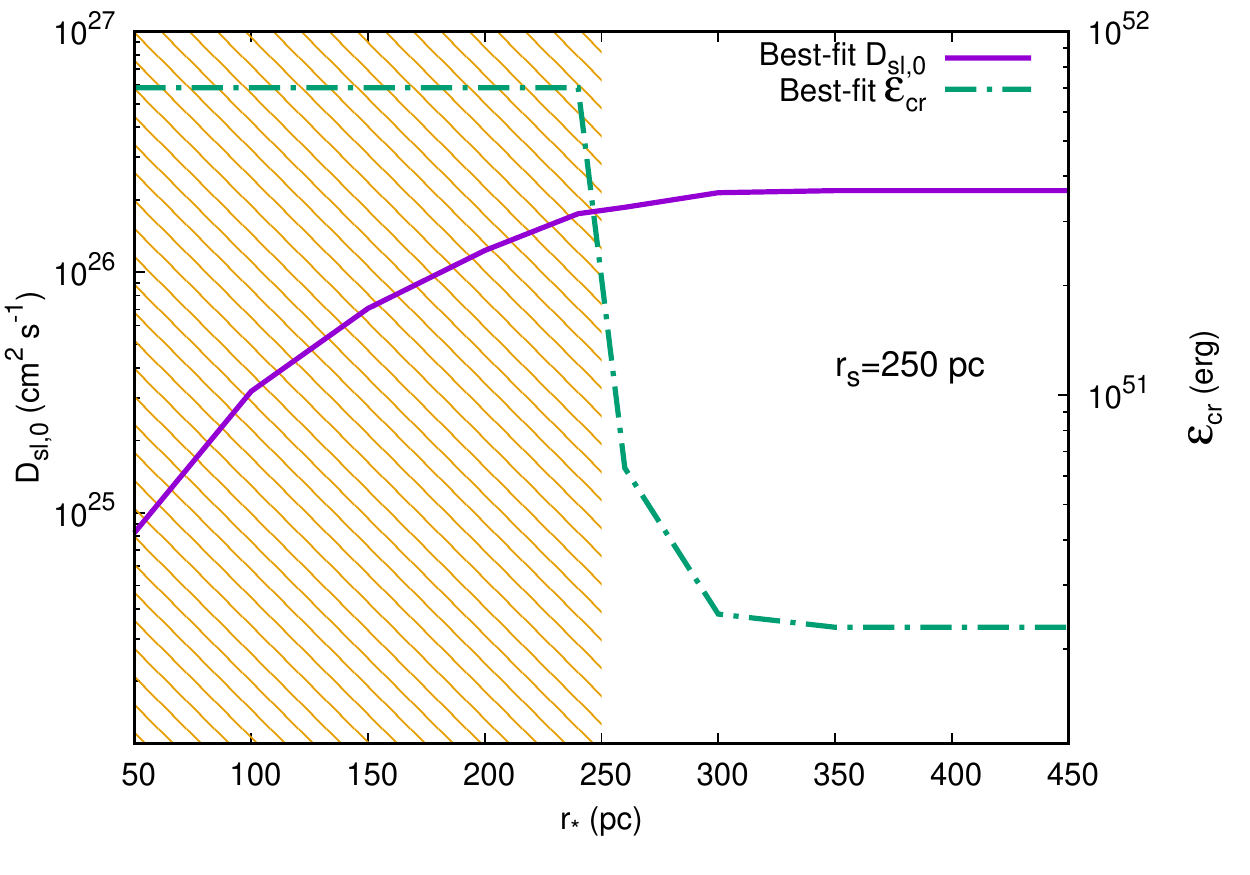}
 \includegraphics[width=0.45\textwidth]{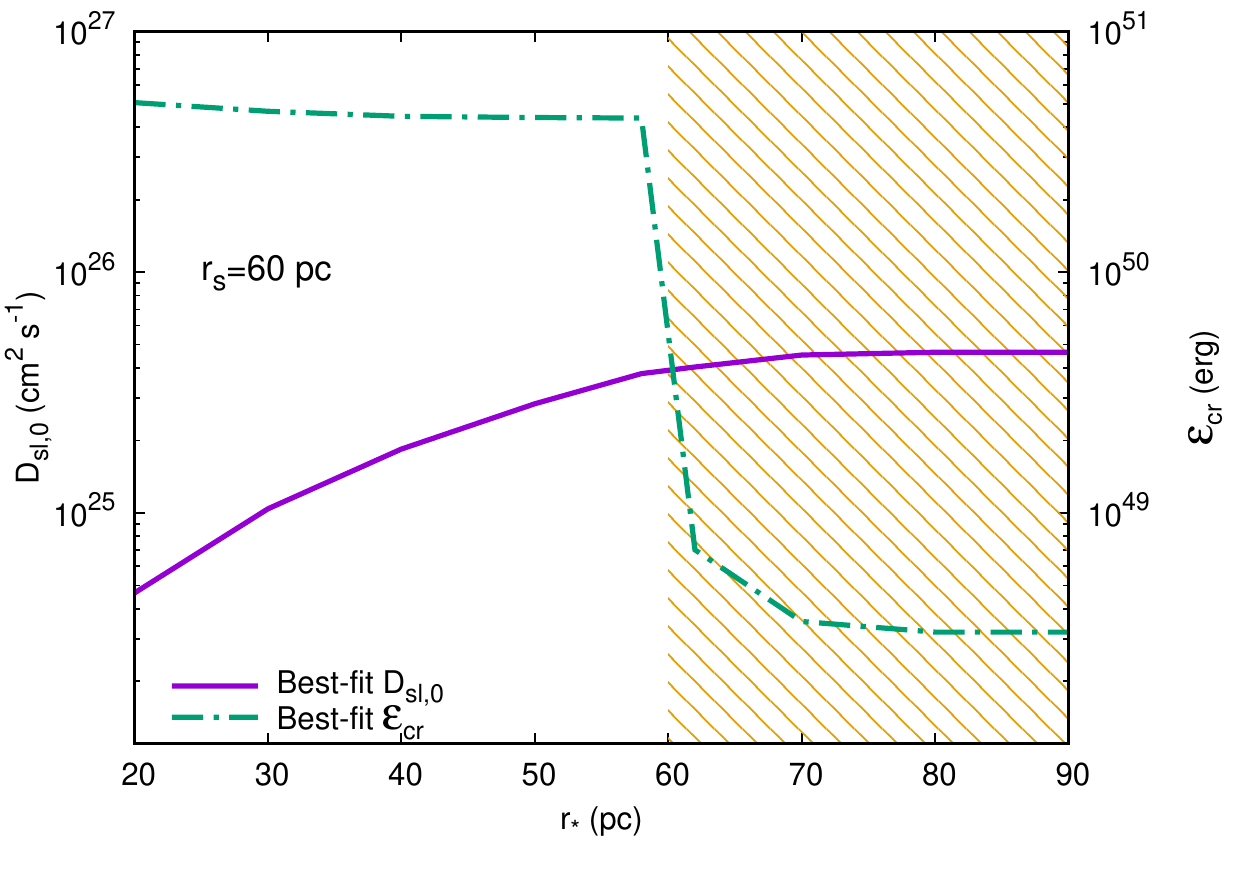}
 \caption{Two possibilities for the slow-diffusion distribution. The abscissa is the slow-diffusion scale centered at the nearby SNR. The unfilled regions show the permitted ranges for the different cases. \textit{Left}: the case of $r_s$=250 pc, which is representative for 100 pc $\lesssim r_s \lesssim$ 500 pc. The slow-diffusion zone must be larger than the distance between the SNR and the solar system, or the CR energy required by the data fitting will be unreasonably large as shown by the shaded area. In this case, the solar system should be embedded in the slow-diffusion region. The homogeneously slow diffusion discussed in Section \ref{sec:value} belongs to this case. \textit{Right}: the case of $r_s=60$ pc, representative for a very nearby SNR. The slow-diffusion scale is constrained to be smaller than $r_s$ in this case.}
 \label{fig:scale}
\end{figure*} 

So far only the anisotropy measurements from Tibet-AS$\gamma$, ARGO-YBJ \citep{2018ApJ...861...93B}, and IceCube \citep{2016ApJ...826..220A} cover the energies both below and above 100 TeV. There is good consistency among all the anisotropy measurements below 100 TeV, while above 100 TeV the result of IceCube is several times lower than the other experiments. We adopt the IceCube data and repeat the fitting procedure in the case of $r_s=250$ pc. The 1D posterior distribution of $D_0$ is shown in the right panel of Fig. \ref{fig:d0}, compared with the case using the Tibet-AS$\gamma$ data. The diffusion coefficient for the IceCube case is a little smaller. As the background anisotropy is required to be lower for the case of IceCube, the background component cancels the nearby SNR component less compared with the case of Tibet-AS$\gamma$. So the anisotropy from the nearby SNR is constrained to be smaller. To maintain the proton spectrum, a smaller diffusion coefficient is then needed for the case of IceCube. In the future, LHASSO will provide more accurate measurements for the CR anisotropy and even aims to give separate measurements for the light and heavy masses \citep{2019arXiv190502773B}.

There is no accurate measurement for the proton spectrum between 100 TeV and 1 PeV at present, so the proton background cannot be well determined. In the above calculations, we assume a broken power-law function for the background and decide the parameters by fitting to the low-energy AMS-02 data and the KASCADE data. The best-fit spectral hardening is $\Delta\alpha\simeq0.086$ as mentioned above. Here we consider the most extreme cases with $\Delta\alpha=0.13$ and $\Delta\alpha=0.01$, which correspond to the upper and lower limits of the KASCADE error band, respectively. In both cases, the fitting procedures indicate that $D_0$ is still constrained in the magnitude of $10^{26}$ cm$^2$ s$^{-1}$, and our conclusion does not change. The future space detector HERD aims to provide precise measurement for CR composition up to PeV \citep{2014SPIE.9144E..0XZ}, which could be decisive for the proton background. 
 
\section{Distribution of the slow diffusion}
\label{sec:scale}
In the previous sections, we assume that the diffusion coefficient is homogeneous in the nearby ISM and find that the local diffusion coefficient must be significantly smaller than the global average in the Galaxy. However, it is also possible that the slow diffusion is only limited in the vicinity of the nearby SNR. CRs released by the SNR may amplify the magnetic field turbulence of the ambient medium through the streaming instability \citep{1971ApJ...170..265S}, which may account for the slow diffusion \citep[e.g.,][]{2013ApJ...768...73M}. So in this section, we assume a two-zone diffusion model to discuss the spatial scale of the slow diffusion, that is, 
\begin{equation}
  D(r)=\left\{
 \begin{aligned}
  & D_{\rm sl},  & r< r_\star \\
  & D_{\rm glob}, & r\geq r_\star\\
 \end{aligned}
 \right.\,,
 \label{eq:2zone}
\end{equation}
where $r$ is the distance from the nearby SNR. Equation (\ref{eq:2zone}) means that the anomalously slow diffusion is only limited inside $r_*$ from the nearby SNR. In the following we test the relation between $r_*$ and the required $D_{\rm sl}$.

First we fix $r_s=250$ pc and perform the calculation as in the previous section again. However, the diffusion coefficient is spatially inhomogeneous, so we adopt the numerical method given by \citet{2018ApJ...863...30F} to solve the propagation equation. The relation between $r_*$ and $D_{{\rm sl},0}$ ($D_{\rm sl}$ at 1 GV) obtained by the fitting procedures is presented in the left panel of Fig. \ref{fig:scale}. Obviously, a slower diffusion speed is required for a smaller slow-diffusion zone to keep the same CR flux at the Earth. When $r_*>r_s$, i.e. the slow-diffusion zone includes the solar system in, we can see that the required diffusion coefficient regresses to the case of homogeneous diffusion discussed in the previous sections. 

We also show the relation between the required CR injection energy and $r_*$ in Fig. \ref{fig:scale}, which provides crucial constraints to the scale of the slow-diffusion zone. The required $\mathcal{E}_{\rm cr}$ has an abrupt decrease at $r_*=r_s$. The reason is that for the two-zone diffusion model, the CR flux inside the slow-diffusion zone is much higher than that outside the zone (see Fig. 1 in \citet{2018ApJ...863...30F}). When $r_*<r_s$, the required injection energy is nearly $10^{52}$ erg, which is unreasonably large for SNRs. While for $r_*>r_s$, $\mathcal{E}_{\rm cr}$ regresses to the typical value as in the homogeneous diffusion scenario. Thus, the slow-diffusion zone should be large enough to include the solar system in. In other words, the solar system should be embedded in a slow-diffusion environment.

At present there is still large uncertainty for the distance of the nearby SNR, so the specific scale of the slow diffusion also cannot be determined. However, as we have discussed in the previous section, the SNR distance should be in a range of 100 pc $\lesssim r_s\lesssim$ 500 pc for the homogeneous diffusion scenario to guarantee a reasonable $\mathcal{E_{\rm cr}}$. So if the SNR distance is between 100 pc and 500 pc, we must have $r_*>r_s$ or the required CR energy will be too large. In this case, the slow-diffusion zone should be at least several hundred parsecs, which is too large to be interpreted by the CR self-excited scenario mentioned above. Such a giant slow-diffusion zone is very likely to be originated from consecutive SNe or collective stellar winds, as stellar feedback is the main source for ISM turbulence \citep{2020PPCF...62a4014F}. Meanwhile, the solar system is known to be embedded in the Local Bubble, which is a giant cavity with high temperature and low density. The average scale of the Local Bubble is larger than 100 pc. It is also believed to be originated from stellar winds or SNe \citep{2018arXiv180106223F}. So the nearby slow-diffusion zone and the Local Bubble may share the same origin, while a further discussion about the link between them is beyond the scope of the present paper.

Beside the scenario of a large slow-diffusion zone, there is an additional possibility. We have discussed the case of 100 pc $\lesssim r_s\lesssim$ 500 pc above. If the SNR distance is further smaller, the required CR energy will be too small for $r_*>r_s$, while it will gradually become reasonable for $r_*<r_s$. We show the case of $r_s=60$ pc in the right panel of Fig. \ref{fig:scale}. The solar system is no longer required to be inside a slow-diffusion region. However, the slow-diffusion zone around the SNR should not be smaller than $\approx20$ pc, or the required diffusion coefficient will be smaller than the Bohm value (for the case of 10 TeV proton and a magnetic field of 3 $\mu$G), which is the lower limit of CR diffusion coefficient. 

With the current measurements, we may not distinguish between the case of a relatively distant SNR with a large slow-diffusion zone pervading the local ISM and the case of a very nearby SNR with a small slow-diffusion zone just around the SNR. If the distance or the age of this nearby CR source can be estimated by a separate method, we may give further judgment about the spatial distribution of the slow diffusion.
 
\section{Conclusion}
\label{sec:conclude}
We provide a new method to study the CR diffusion coefficient in the nearby ISM. We interpret the DAMPE proton spectrum with a nearby SNR superposed on a continuous proton background. The proton spectrum from the nearby SNR is required to be very hard below $\approx10$ TeV to accommodate the DAMPE data. As the dipole anisotropy measurements of CRs give a constraint to the distance-to-age ratio of the nearby CR source, we illustrate that the DAMPE spectrum cannot be reproduced by simply varying the distance or the age of the SNR. A diffusion coefficient significantly smaller than the Galactic average is necessary in the nearby ISM to reconcile the proton spectrum and the anisotropy measurement. A harder CR injection spectrum for the nearby SNR also cannot reasonably solve this problem without significantly reducing the diffusion coefficient.

Assuming that the property of the local ISM is homogeneous, we fit the proton spectra  and the CR anisotropy to constrain the diffusion coefficient. The required $D_0$ is limited in the magnitude of $10^{26}$ cm$^2$ s$^{-1}$, which is about 100 times smaller than the global average derived with B/C. Meanwhile, $D_0$ around two nearby pulsars measured by HAWC is also $\approx10^{26}$ cm$^2$ s$^{-1}$. The slow diffusion observed by HAWC should indeed happen in the ISM, rather than be interpreted by confined structures like pulsar wind nebulae \citep{2019arXiv190712121G}. Thus, the two separate methods reach the same conclusion about the local CR diffusion. 

We further discuss the spatial distribution of the slow diffusion by a two-zone diffusion model. If the SNR distance is several hundred of parsecs, the slow-diffusion zone is required to be large enough to include both the SNR and the solar system in, to guarantee a reasonable CR injection energy. The homogeneous diffusion scenario mentioned above belongs to this case. Such a large slow-diffusion zone is likely to be originated from consecutive SNe or collective stellar winds. There is another possibility that the source is very nearby with a distance of tens of parsecs. In this case the slow-diffusion is limited in the vicinity of the SNR.

As a by product, the information of the nearby SNR is constrained to some degree. The distance of the SNR should be smaller than 500 pc, while the age should be in the range of 0.1-1 Myr.
 
\section*{Acknowledgement}
This work is supported by the National Key Program for Research and Development 
(No.~2016YFA0400200) and by the National Natural Science 
Foundation of China under Grants No.~U1738209,~11851303.


\end{document}